\documentclass[12pt]{article}
\usepackage{a4wide}
\usepackage{amssymb}
\usepackage{graphicx}
\begin{document}
{\renewcommand{\thefootnote}{\fnsymbol{footnote}}
\begin{center}
{\LARGE  Inequivalence of mimetic gravity\\[2mm] with models of loop quantum gravity}\\
\vspace{1.5em}
Martin Bojowald\footnote{e-mail address: {\tt bojowald@psu.edu}}
and Erick I.~Duque\footnote{e-mail address: {\tt eqd5272@psu.edu}}
\\
\vspace{0.5em}
Institute for Gravitation and the Cosmos,\\
The Pennsylvania State
University,\\
104 Davey Lab, University Park, PA 16802, USA\\
\vspace{1.5em}
\end{center}
}

\setcounter{footnote}{0}

\begin{abstract}
  Certain versions of mimetic gravity have recently been claimed to present
  potential covariant theories of canonically modified spherically symmetric
  gravity, motivated by ingredients from loop quantum gravity. If such an
  equivalence were to hold, it would demonstrate general covariance of a large
  class of models considered in loop quantum gravity. However, the
  relationship with mimetic gravity as presented so far is incomplete because
  it has been proposed only in preferred space-time slicings of uniform scalar
  fields. Here, several independent arguments are used to show that neither an
  equivalence nor a covariance claim are correct for models of loop quantum
  gravity. The framework of emergent modified gravity is found to present a
  broad setting in which such questions can be analyzed efficiently. As an
  additional result, the discussion sheds light on the co-existence of
  different and mutually inequivalent approaches to an implementation of the
  gravitational dynamics within loop quantum gravity.
\end{abstract}

\section{Introduction}

Canonical quantum gravity is expected to change the structure of space-time by
incorporating general quantum effects such as fluctuations or
discreteness. The dynamics of general relativity, no longer unfolding on a
continuous Riemannian geometry, is then likely to be modified, giving rise to
potential new phenomena in particular at high curvature, such as at the big
bang or in black holes. However, the definition of physical effects relevant
for the expanding universe or horizons requires certain geometrical notions
that work well in the classical continuum but become less clear, at least
according to the present stage of knowledge, in fundamentally discrete
theories. Investigations of potential physical effects of proposals for
quantum gravity therefore aim to strike a balance between including
sufficiently many interesting dynamical modifications while trying to stay
close enough to the classical space-time structure in order to apply familiar
notions of curvature scalars or geodesic properties. An important tool in such
effective descriptions is given by space-time metric tensors or line elements,
suitably modified by supposed quantum effects.

While effective line elements can be used to apply classical definitions to
proposed quantum solutions, they are not always available because they
implicitly assume that expressions used for modified metric tensors indeed
enjoy proper tensor transformations under coordinate changes. This condition
ensures that the line element is, as required, invariant under coordinate
transformations and therefore presents an unambiguous description of
space-time geometry. In models of canonical quantum gravity, however, the
underlying equations for gauge transformations that classically are equivalent
to coordinate changes are usually modified. It is then not guaranteed that
modified metric components obey the condition required for a well-defined line
element. In more general terms, it is far from clear whether canonical quantum
gravity or its effective models can be consistent with general covariance at
least in regimes in which one tries to apply the usual notions of curvature
scalars or horizons that are based on the existence of an invariant line
element.

In models of loop quantum gravity, the covariance question has been analyzed
in some detail, resulting in several no-go results for previously proposed
formulations \cite{SphSymmCov,GowdyCov,NonCovDressed,NonCovPol,Disfig}. It is
therefore surprising that a recent analysis \cite{MimeticCGHS} concludes that
some of the models that had already been ruled out as non-covariant or even
anomalous may be described in an equivalent way by certain versions of mimetic
gravity, which are generally covariant. However, the arguments presented there
are incomplete because they demonstrate a relationship between canonical
equations in models of loop quantum gravity and tensor equations derived from
mimetic gravity only in a preferred set of slicings. This set of slicings is
defined by uniform scalar fields that by definition occur in mimetic gravity,
but are included by hand on the canonical side in order to perform simple
deparameterizations of the constrained dynamics. The relationship can be
extended to the general theories, then holding for any slicing, only if one of
the following two assumptions is met: (i) Gauge transformations of the
modified canonical theory are compatible with slicing changes or coordinate
transformations in space-time, or (ii) the mimetic version reduced to
alternative gauge choices with non-uniform scalar field in some way resembles
Hamiltonians expected from the procedures of loop quantum gravity. The first
assumption is precisely the property that had already been ruled out for some
of the same models, for instance in \cite{SphSymmCov}, without using the
connection to mimetic gravity. The second assumption has been made implicitly
but not evaluated further in \cite{MimeticCGHS}.

We will show that the claimed equivalence of mimetic gravity and models of
loop quantum gravity, as well as general covariance of the latter, can quickly
and easily be disproved because the same constructions could be applied to
versions of Ho\v{r}ava--Lifshitz gravity \cite{Horava} that are known not to
be covariant and are physically inequivalent to mimetic gravity. Nevertheless,
given the importance of the covariance question in models of canonical and
loop quantum gravity, it is of interest to provide a detailed analysis in
order to determine where exactly the equivalence fails, tracing it back to
specific terms in the Hamiltonian constraint of spherically symmetric general
relativity and its possible modifications. We will present these details in
Section~\ref{s:Covariance}. In Section~\ref{s:Emergent}, we will perform
a related analysis within the framework of emergent modified gravity
\cite{Higher,HigherCov} in which it is possible to derive all compatible
covariant modifications up to a given order in derivatives. Before our
Conclusions, Section~\ref{s:Approaches} places our results in the context of
different approaches within loop quantum gravity.

\section{Covariance within mimetic gravity}
\label{s:Covariance}

The constructions in \cite{MimeticCGHS} make use of the Lagrangian
\begin{equation}
  S[g,\phi,\lambda]=\frac{1}{16\pi G} \int{\rm d}^4x \sqrt{-\det
    g}\left(R+ L[\phi]+
    \lambda((\nabla_{\mu}\phi)(\nabla^{\mu}\phi)+1)\right)
    \label{eq:Mimetic action}
\end{equation}
where
\begin{equation} \label{Lphi}
  L[\phi]=
  L(\nabla^{\mu}\nabla_{\mu}\phi,(\nabla_{\mu}\nabla_{\nu}\phi)(\nabla^{\mu}\nabla^{\nu}\phi))
\end{equation}
and $\lambda$ is a Lagrange multiplier, as a special version of mimetic
gravity \cite{Mimetic,MimeticRev,MimeticDHOST}. Mimetic gravity had previously
been used for models of loop quantum cosmology \cite{LimCurvLQC}, where it had
been found not to provide an equivalence \cite{MimeticLQC,MimeticLQCPert} if
more complicated features beyond purely isotropic models were included. It
would therefore be surprising if mimetic gravity could be used as a complete
equivalence for general spherically symmetric models, as claimed in
\cite{MimeticCGHS}.

The 4-dimensional action is then reduced to a canonical formulation of
spherically symmetric configurations by assuming a space-time metric according
to the line element
\begin{equation}
  {\rm d}s^2=-N^2{\rm d}t^2+ \frac{(E^{\varphi})^2}{E^x}({\rm d}x+M{\rm
    d}t)^2+ E^x({\rm d}\vartheta^2+\sin^2\vartheta{\rm d}\varphi^2)
\end{equation}
with the lapse function $N$, the shift vector $M$, and a spatial metric
expressed in terms of densitized-triad components $E^{\varphi}$ and $E^x$
(assumed positive, fixing the spatial orientation) as used in models of loop
quantum gravity \cite{SymmRed,SphSymm,SphSymmHam}. The scalar field and
Lagrange multiplier are assumed to depend only on $t$ and $x$ in order to
respect spherical symmetry. The resulting scalar-tensor theory is covariant
under coordinate transformations or slicing changes that preserve spherical
symmetry.

\subsection{Gauge fixing and Legendre transformation}

In the next step of \cite{MimeticCGHS}, gauge-fixing conditions are
introduced. A preferred slicing is defined by uniform scalar fields, such that
$\phi(t)$ is no longer allowed to depend on $x$. Locally, this choice is
always possible in a covariant theory. The mimetic condition
\begin{equation}
  0=g^{\mu\nu}(\nabla_{\mu}\phi)(\nabla_{\nu}\phi)+1=-\frac{\dot{\phi}^2}{N^2}+1
  \label{eq:Mimetic condition}
\end{equation}
then determines the lapse function, which also depends only on $t$. Locally,
one may choose $t=\phi$ such that $N=1$. The scalar terms in the reduced
action can then be evaluated explicitly in terms of derivatives of the triad
components, for which we will provide more details in
Section~\ref{s:Approaches}. The result is that a generic spherically symmetric
gauge-fixed action takes the form
\begin{equation} \label{S}
  S=\frac{1}{2G} \int{\rm d}t{\rm d}x
  NE^{\varphi}\sqrt{E^x}\left(-2XY+Y^2+\tilde{L}(X,Y)
    +\frac{1}{2}R^{(3)}\right)
\end{equation}
where $R^{(3)}$ is the spatial Ricci scalar,
\begin{equation}
  R^{(3)}= \frac{(E^x)'(E^{\varphi})'}{(E^{\varphi})^3}-
  \frac{((E^x)')^2}{4E^x(E^{\varphi})^2}- \frac{(E^x)''}{(E^{\varphi})^2}+
  \frac{1}{E^x}\,,
\end{equation}
and we have
\begin{equation} \label{XY}
  X=\frac{\dot{E}^{\varphi}-(ME^{\varphi})'}{NE^{\varphi}}\quad,\quad
  Y=\frac{\dot{E}^x-M(E^x)'}{2NE^x}
  \,.
\end{equation}

The function $\tilde{L}$ is determined by the original function $L$ of mimetic
gravity. The construction relies on a coincidence in the gauge used, which
implies that extrinsic-curvature components of a spherically symmetric metric,
which imply the terms $-2XY+Y^2$ in the action using Gauss--Codazzi
relationships, can be expressed uniquely through the same two functions, $X$
and $Y$, that determine the independent scalar contributions to a spherically
symmetric mimetic theory with a dependence of the form (\ref{Lphi}). We will
return to this observation in Section~\ref{s:Approaches}.

There is also a class of deformations in \cite{MimeticCGHS} parameterized by a
real number $\eta$ for purposes of realizing different models of dilaton
gravity with actions
\begin{equation} \label{Seta}
  S_{\eta}=\frac{1}{2G} \int{\rm d}t{\rm d}x
  NE^{\varphi}\sqrt{E^x}\left(-2XY+(1-\eta) Y^2+\tilde{L}(X,Y)
    +\frac{1}{2}R^{(3)}_{\eta}\right)
\end{equation}
where
\begin{equation} \label{R3eta}
  R^{(3)}_{\eta}= \frac{(E^x)'(E^{\varphi})'}{(E^{\varphi})^3}-
  (1-\eta)\frac{((E^x)')^2}{4E^x(E^{\varphi})^2}- \frac{(E^x)''}{(E^{\varphi})^2}+
  \frac{\eta+1}{(E^x)^{1-\eta}}\,.
\end{equation}
Although we will not be interested in dilaton gravity in the present paper, it
is useful to keep the parameter $\eta$ for our covariance analysis.

The action functional determines momenta
\begin{equation}
  \pi_{\varphi}= 2G\frac{\delta S_{\eta}}{\delta\dot{E}^{\varphi}}=
  \sqrt{E^x}(\partial_X\tilde{L}-2Y)
\end{equation}
and
\begin{equation}
  \pi_x=2G\frac{\delta
    S_{\eta}}{\delta\dot{E}^x}=\frac{E^{\varphi}}{2\sqrt{E^x}}
  (\partial_Y\tilde{L}-2X+2(1-\eta)Y)
\end{equation}
canonically conjugate to $E^{\varphi}$ and $E^x$, respectively. A Legendre
transformation leads to the Hamiltonian
\begin{eqnarray}
  H_{\eta}&=& -\frac{1}{2G} \int{\rm d}x N\sqrt{E^x}E^{\varphi}
  \left(\tilde{L}-X\partial_X\tilde{L}-Y\partial_Y\tilde{L} +2XY-
              (1-\eta)Y^2+\frac{1}{2}R^{(3)}_{\eta}\right)\nonumber\\
  &&+\frac{1}{2G}\int{\rm d}x
  (M(E^x)'\pi_x+(ME^{\varphi})'\pi_{\varphi})\,.
  \label{eq:Hamiltonian gauge-fixed}
\end{eqnarray}
Varying by $M$, the second line implies the diffeomorphism constraint, and if
$N$ had not been gauge fixed, the first line would imply the Hamiltonian
constraint.

The case of $\tilde{L}=0$ should correspond to vacuum spherically
symmetric gravity. A direct calculation indeed yields
\begin{equation}
  Y_{\tilde{L}=0}=-\frac{\pi_{\varphi}}{2\sqrt{E^x}} \quad,\quad
  X_{\tilde{L}=0}= -\frac{\sqrt{E^x}}{E^{\varphi}} \pi_x-
  \frac{1-\eta}{2\sqrt{E^x}} \pi_{\varphi}
\end{equation}
and therefore
\begin{equation} \label{XYclass}
  2X_{\tilde{L}=0}Y_{\tilde{L}=0}- (1-\eta)Y_{\tilde{L}=0}^2=
  \frac{\pi_{\varphi}\pi_x}{E^{\varphi}}+ \frac{1-\eta}{4E^x}\pi_{\varphi}^2
\end{equation}
which produces the correct momentum-dependent terms for the Hamiltonian
constraint used, for instance, in \cite{HigherCov} if $\eta=0$.

\subsection{Modifications and violations of covariance}
\label{s:Mod}

Continuing for now with $\tilde{L}=0$, the reduced gauge-fixed theory is
equivalent to vacuum spherically symmetric general relativity and therefore
covariant. The procedure outlined in \cite{MimeticCGHS} then attempts to map
modifications of the canonical theory, such as periodic dependences on the
momenta as motivated by loop quantum gravity, to suitable non-zero
$\tilde{L}$. In this way, models of loop quantum gravity are mapped to
specific versions of gauge-fixed action principles for spherically symmetric
configurations. Without further discussion, it is then claimed that the
original theories must be covariant because they are strictly related to
mimetic gravity, which is covariant.

However, taken at face value, the correspondence only shows that models of
loop quantum gravity are equivalent to certain gauge-fixed action principles
for spherically symmetric configurations. It does not follow that the gauge
fixing can be relaxed while maintaining the correspondence. To be sure, there
are scalar fields on both sides, used for deparameterization in models of loop
quantum gravity and included as dynamical matter on the mimetic side. But the
specific contributions of scalar fields to constraints or actions are never
compared with each other. It is therefore unclear whether the specific
dynamical dependence of the mimetic action on $\phi$ via $L$, whose tensorial
nature is important for general covariance, is equivalent to the form in which
the scalar field is implemented on the canonical side for models of loop
quantum gravity. Without careful adjustments, which have not been made when
deparameterizing the models, it is highly unlikely that the scalar terms match
and complete all conditions for general covariance. In an alternative
viewpoint, one may use the covariant mimetic theory in order to define what
the scalar couplings of the canonical theory should look like in any gauge for
it to be covariant. The main question, whether the required scalar terms have
a good chance of resembling what is usually considered a model of loop quantum
gravity, will be discussed in Section~\ref{s:Approaches}.

On the mimetic side, there are two versions of canonical spherically symmetric
reductions, first the symmetry reduction by itself and then the reduction on a
preferred slicing given by uniform $\phi$. For simplicity, we will call the
former the $(E,\pi,\phi)$-theories and the latter the $(E,\pi)$-theories (both
for various $\tilde{L}$) since in this case $\phi$ has been eliminated by
gauge fixing. While the $(E,\pi,\phi)$-theories are clearly covariant, as
symmetry reductions of 4-dimensional covariant theories, the correspondence with
models of loop quantum cosmology envisaged in \cite{MimeticCGHS} only
considers the $(E,\pi)$-theories. We will first see whether a covariance
argument in the sense of slicing independence can be made for these theories
with generic $\tilde{L}$, for which the equivalence to spherically symmetric
gravity is not available.

Starting with a spatial slice $\Sigma$ with initial values $E_0$, $\pi_0$ and
$\phi_0$ of an $(E,\pi)$-theory, it can always be embedded in a corresponding
$(E,\pi,\phi)$-theory with the same $\tilde{L}$ in which $\Sigma$ is realized
as a spatial slice $\Sigma'\colon\phi=\phi_0$ with $\phi_0$ constant. On this
slice, we may impose the same initial values $E_0$ and $\pi_0$ for the
gravitational fields and choose the momentum of $\phi$ such that the
Hamiltonian constraint is satisfied. (The diffeomorphism constraint does not
depend on the momentum since $\phi=\phi_0$ is constant on the initial slice,
such that $\phi'\pi_{\phi}=0$.)  Since all $(E,\pi,\phi)$-theories are
covariant, we know that its solutions determine space-time geometries, such
that there are transformations to a different slice $\Sigma''$ on which we
have new values $E_1$, $\pi_1$, $\phi_1$ and scalar momenta obtained from the
original values on $\Sigma'$. For a covariance argument, we would then like to
restrict field values on $\Sigma'$ to just the gravitational fields and
conclude that the transformation could be interpreted as a slicing change in
the $(E,\pi)$-theory. However, this is not possible because there is no
guarantee that the new field $\phi_1$ on $\Sigma''$ is constant, as required
by definition of the gauge-fixed $(E,\pi)$-theory.

In fact, it can be shown explicitly that, in general, $\phi$ cannot be
constant on a new slice obtained by a coordinate transformation.  Consider a
general coordinate transformation $(t,x) \to (\tilde{t}, \tilde{x})$, with new
coordinates indicated by a tilde.  If we express the original coordinates (in
which $\phi'=\partial\phi/\partial x=0$ on the given slice), as function of the new
ones, $t (\tilde{t} , \tilde{x})$ and $x (\tilde{t} , \tilde{x})$, we have
\begin{eqnarray}
    \frac{\partial \phi}{\partial \tilde{x}} = \dot{\phi} \frac{\partial t}{\partial \tilde{x}}
    + \phi' \frac{\partial x}{\partial \tilde{x}}
    = \dot{\phi} \frac{\partial t}{\partial \tilde{x}}
    \ .
\end{eqnarray}
If $\partial t / \partial \tilde{x}\neq0$ then in the new coordinate system we
have $\tilde{\phi'}\neq 0$ (taking the spatial derivative in the new
coordinate system, as referred to by the tilde). Therefore, this gauge is not
contained in the $(E,\pi)$-theories, in contrast to what is implicitly assumed
by using covariance statements for the gauge-fixed mimetic theory. In the new
coordinate system, the Hamiltonian (\ref{eq:Hamiltonian gauge-fixed}) for a
general function $\tilde{L}$ is no longer recovered because the latter
descends from the $\nabla_\mu \phi$-dependent $L$ in (\ref{eq:Mimetic
  action}), and extra terms appear in the new gauge if $\tilde{L}\not=0$. (See
equations~(\ref{fullY}) and (\ref{fullX}) to be discussed in
Section~\ref{s:Approaches}.)

We conclude that slicing independence or covariance of mimetic gravity and its
spherically symmetric reduction does not imply slicing independence of the
gauge-fixed theory which is formulated only for the gravitational fields. For
some $\tilde{L}$, the gauge-fixed theory may be covariant as in the case of
$\tilde{L}=0$, but checking covariance for non-zero $\tilde{L}$ requires
additional conditions that have not been considered in \cite{MimeticCGHS} for
models of loop quantum gravity. We will fill this gap in the remainder of this
paper, showing that covariance is, in fact, violated.

\subsection{Quadratic theories and Ho\v{r}ava--Lifshitz gravity}

A function $\tilde{L}(X,Y)=aX^2+bXY+cY^2$ can be used to generate an arbitrary
quadratic dependence of the Hamiltonian on momenta. To see this, we simply
evaluate the contribution
\begin{eqnarray}
  &&\tilde{L}-X\partial_X\tilde{L}-Y\partial_Y\tilde{L}+2XY-(1-\eta)Y^2\nonumber\\
  &=&
                                                                         -aX^2+(2-b)XY-
                                                                         (1-\eta+c)Y^2\\
  &=& -a\frac{E^x}{(E^{\varphi})^2} \pi_x^2+ (1-b/2-a(1-\eta))
      \frac{\pi_x\pi_{\varphi}}{E^{\varphi}} +((1-b)(1-\eta)-c-a(1-\eta)^2)
      \frac{\pi_{\varphi}^2}{4E^x}\,.\nonumber
\end{eqnarray}
For suitable $a$, $b$ and $c$ (irrespective of $\eta$), any quadratic
dependence on the momenta can be generated.

However, it is well known that the dependence of the Hamiltonian on momenta or
of the action on extrinsic curvature is restricted by covariance. Quadratic
momenta without higher time derivatives imply classical theories. Without
symmetry reduction, the quadratic dependence of the action on extrinsic
curvature $K_{ab}$ is completely determined as $K^{ab}K_{ab}-(K^a_a)^2$, which
in a spherically symmetric reduction implies the specific coefficients of
$\pi_{\varphi}^2$, $\pi_{\varphi}\pi_x$ and $\pi_x^2$ seen in
(\ref{XYclass}): In a triad basis, $K_{ab}$ has the components $K_x$ and
$K_{\varphi}$ where $K_x=\pi_x$ and $2K_{\varphi}=\pi_{\varphi}$
\cite{SphSymm}. Therefore, supplying density weights by using the spatial metric $q_{ab}$,
\begin{equation}
  \sqrt{\det q}K^a_a=K_a^iE^a_i=K_xE^x+2K_{\varphi}E^{\varphi}
\end{equation}
and
\begin{equation}
  \sqrt{\det q} K^{ab}K_{ab}= \frac{1}{\sqrt{\det q}} K_a^iE^b_i K_b^jE_j^a=
  \frac{(E^x)^{3/2}}{E^{\varphi}} K_x^2+2\frac{E^{\varphi}}{\sqrt{E^x}}
  K_{\varphi}^2\,.
\end{equation}
The combination
\begin{equation}
  \sqrt{\det q} (K^{ab}K_{ab}-(K^a_a)^2)=
  -4\sqrt{E^x}K_xK_{\varphi}-2\frac{E^{\varphi}}{\sqrt{E^x}} K_{\varphi}^2
\end{equation}
is indeed proportional to (\ref{XYclass}). The possibility of using an unrestricted quadratic
$\tilde{L}$ in gauge-fixed reduced theories of mimetic gravity does not respect
the required conditions on coefficients.

It is in fact possible to generalize spherically symmetric theories while
keeping them covariant, as seen in 2-dimensional dilaton gravity. The
coefficients in (\ref{XYclass}) are unique for the reduction of general
relativity to spherical symmetry. They may be changed within 2-dimensional
covariant theories, but still not arbitrarily so, as it would be suggested by a general
quadratic $\tilde{L}$. We will discuss restrictions on covariant 2-dimensional
theories in the next section, and now show that the generality of quadratic
$\tilde{L}$ can be used to construct a correspondence between gauge-fixed
reduced mimetic gravity and Ha\v{r}ava--Lifshitz theories that are known not
to be covariant.

In 4-dimensional space-time, a class of Ho\v{r}ava--Lifshitz gravity theories
\cite{Horava} is defined by the action
\begin{equation}
  S_{\rm HL}^{\lambda}=\frac{1}{16\pi G} \int{\rm d}^4x N\sqrt{\det
    q}\left(K^{ab}K_{ab}-\lambda(K^a_a)^2 +\frac{1}{1-3\lambda} R^{(3)}\right)
\end{equation}
where $q_{ab}$ and $K_{ab}$ are metric and extrinsic-curvature tensors defined
on the slices of a foliation. For $\lambda=1$, the theory is a restriction of
general relativity to the foliation, but for $\lambda\not=1$ there is no
such correspondence to a covariant theory. If we repeat our reduction of the
$K$-terms to spherical symmetry, we obtain
\begin{equation} \label{KHL}
  \sqrt{\det q} (K^{ab}K_{ab}-\lambda (K^a_a)^2)=
  (1-\lambda) \frac{(E^x)^{3/2}}{E^{\varphi}}
  K_x^2-4\lambda\sqrt{E^x}K_xK_{\varphi}+2(1-2\lambda)\frac{E^{\varphi}}{\sqrt{E^x}}
  K_{\varphi}^2 \,.
\end{equation}
For every $\lambda$, there is a $\tilde{L}$ such that this combination of
$K_x$ and $K_{\varphi}$ has a correspondence with gauge-fixed reduced mimetic
gravity according to \cite{MimeticCGHS}. Since generic Ho\v{r}ava--Lifshitz
theories are not covariant, this correspondence cannot be used to prove
covariance in other cases, such as models of loop quantum gravity. Moreover, the
correspondence does not imply physical equivalence of the formally
related theories.

If we consider only 2-dimensional theories of Ho\v{r}ava--Lifshitz type, there
is no need to use the reduction (\ref{KHL}). A general 2-dimensional theory of
this form can then be defined with $K$-terms proportional to
\begin{equation} \label{HLKsphsymm}
  K_{\varphi}^2 +\lambda_1 \frac{K_xK_{\varphi}}{E^{\varphi}}+
  \lambda_2\frac{K_x^2}{(E^{\varphi})^2}
\end{equation}
with two parameters, $\lambda_1$ and $\lambda_2$. In spherical symmetry, there
are three independent spatial scalars quadratic in the momenta, constructed by
referring to the density weight zero for $K_{\varphi}$ and $E^x$ and density
weight one for $K_x$ and $E^{\varphi}$. Gauge-fixed reduced mimetic theories
with quadratic $\tilde{L}$ are sufficiently general to include even these
theories. There are no restrictions that could suggest any potential
implementation of conditions for general covariance through the
correspondence.

\section{Covariance from emergent modified gravity}
\label{s:Emergent}

We have seen that the gauge-fixed reduced theory of mimetic gravity with
action (\ref{S}) is covariant if $\tilde{L}=0$ because it then equals a gauge
fixing of vacuum spherically symmetric gravity, which clearly has a covariant
extension to arbitrary slicings. This conclusion remains true if we use
(\ref{Seta}) with arbitrary $\eta$ because $\tilde{L}=0$ implies that the
scalar field does not couple to gravity at all; it is merely used to define a
slicing. If the scalar field does not couple to gravity, the generic
hypersurface deformations discussed in Section~\ref{s:Mod} can always be
performed such that $\phi$ remains unchanged, and in particular constant, when
transforming to a new slicing. Slicings of an $(E,\pi)$-theory can then always
be embedded in slicings of an $(E,\pi,\phi)$-theory, which are covariant. The
action (\ref{Seta}) with $\tilde{L}=0$ therefore defines a covariant
2-dimensional theory even though, for non-zero $\eta$, it modifies the
classical relationship between coefficients of the momentum terms
(\ref{XYclass}) as well as the spatial Ricci scalar (\ref{R3eta}).

We are again seeing an example of the fact that 2-dimensional covariant models
allow for more freedom than spherically symmetric reductions of 4-dimensional
covariant theories. According to  \cite{Action}, any Hamiltonian constraint
of the form
\begin{equation} \label{Halpha}
  H=-\frac{1}{2G} \int{\rm d}xN\left(\alpha\frac{E^{\varphi}}{\sqrt{E^x}}
    K_{\varphi}^2+ 2\bar{\alpha}\sqrt{E^x}K_xK_{\varphi}+ \alpha_{\Gamma}
    \frac{E^{\varphi}}{\sqrt{E^x}} (1-\Gamma_{\varphi})^2+
    2\bar{\alpha}_{\Gamma} \sqrt{E^x}\Gamma_{\varphi}'\right)
\end{equation}
with
\begin{equation}
  \Gamma_{\varphi}= -\frac{(E^x)'}{2E^{\varphi}}
\end{equation}
defines an anomaly-free canonical theory in two dimensions, provided the
$E^x$-dependent functions $\alpha$, $\bar{\alpha}$, $\alpha_{\Gamma}$ and
$\bar{\alpha}_{\Gamma}$ obey
\begin{equation}
  (\bar{\alpha}\alpha_{\Gamma}-\alpha\bar{\alpha}_{\Gamma})(E^x)'+
  2(\bar{\alpha}'\bar{\alpha}_{\Gamma}
  -\bar{\alpha}\bar{\alpha}_{\Gamma}')E^x=0\,.
\end{equation}
This equation can be solved by the parameterization
\begin{eqnarray}
  \bar{\alpha} = \sqrt{|\beta|}b_1\quad&,&\quad \alpha=\sqrt{|\beta|}b_1b_2\\
  \bar{\alpha}_{\Gamma}= {\rm
  sgn}(\beta)\frac{\sqrt{|\beta|}}{b_1}\quad&,&\quad \alpha_{\Gamma}={\rm
                                                sgn}(\beta)
                                                \frac{\sqrt{|\beta|}}{b_1}\left(b_2-4\frac{{\rm
                                                d}\log b_1}{{\rm d}\log
                                                E^x}\right)
\end{eqnarray}
in terms of three functions, $b_1$, $b_2$ and $\beta$. The structure function
in the Poisson bracket of two Hamiltonian constraints is then multiplied by
$\beta$. Recently, these theories have been shown to be not only anomaly-free
but also generally covariant \cite{HigherCov} provided the radial component of
the spatial metric of a compatible space-time line element is given by
$|\beta|^{-1}(E^{\varphi})^2/E^x$ rather than the classical
$(E^{\varphi})^2/E^x$. The signature of the space-time metric is determined by
${\rm sgn}(\beta)$, being Lorentzian if $\beta>0$ and Euclidean if $\beta<0$
\cite{EffLine}. Moreover, a term of the form $NE^{\varphi}V(E^x)$ with an
arbitrary dilaton potential $V(E^x)$ can be added to the integrand of the
Hamiltonian constraint (\ref{Halpha}) without introducing anomalies or
violating covariance.

There is therefore a large class of covariant 2-dimensional theories that do
not have extensions to covariant 4-dimensional theories, all quadratic in
momenta. The canonical theory defined by (\ref{Seta}) with (\ref{R3eta}) is a
simple example of these models in which $\bar{\alpha}=1=\bar{\alpha}_{\Gamma}$
and $\alpha=\alpha_{\gamma}=1-\eta$ are constant, while the dilaton potential
$V(E^x)$ can be used to account for the power $(E^x)^{\eta-1}$ in the last
term of (\ref{R3eta}). This result confirms that any gauge-fixed reduction of
mimetic gravity with $\tilde{L}=0$, eliminating scalar couplings to gravity,
defines a covariant 2-dimensional theory. In terms of the parameterization by
$\beta$, $b_1$ and $b_2$, the condition $\bar{\alpha}=\bar{\alpha}_{\Gamma}$
implies that ${\rm sgn}(\beta)=b_1^2$, or $\beta>0$ and $b_1=1$. Since both
$\bar{\alpha}$ and $\bar{\alpha}_{\Gamma}$ equal one in this case, we obtain
$\beta=1$.  The second condition, $\alpha=\alpha_{\gamma}=1-\eta$, then
equates $b_2$ with the constant $1-\eta$. With $\beta=1$, there is no
signature change and the classical expression $(E^{\varphi})^2/E^x$ can be
used as the radial metric component for all $\eta$. It is important to note
that this result must be derived from the constraint brackets and covariance
conditions. In general, it cannot be taken for granted.

For non-linear $\tilde{L}$, the Hamiltonian contains a term of the form
$K_x^2$ or higher orders, which is not included in (\ref{Halpha}) and cannot
be made compatible with general covariance as shown by the results of emergent
modified gravity. This term in (\ref{HLKsphsymm}) is therefore the crucial
difference between symmetry properties of Ho\v{r}ava--Lifshitz gravity and
dilaton gravity in two dimensions. The attempted relationship with models of
loop quantum gravity constructed in \cite{MimeticCGHS} considers terms of the
form $\sin(\Delta_1 K_x)$ where $\Delta_1$ may depend on $E^x$ and
$E^{\varphi}$. In a Taylor expansion, this function includes $K_x^2$-terms as
well as higher orders in $K_x$, violating covariance.  Terms of the form
$\sin(\Delta_2K_{\varphi})$, again with a possible dependence of $\Delta_2$ on
$E^x$ (but not on $E^{\varphi}$) can be made covariant, but, as shown in
\cite{HigherCov}, only by a canonical transformation that introduces the sine
function by hand. Specific properties of this function, often related to
fundamental properties of loop quantum gravity, therefore cannot have physical
implications on the level of effective space-time geometries. Finally,
covariance of these modifications requires the presence of
$K_{\varphi}'$-terms in the Hamiltonian constraint or coupling terms of
$K_{\varphi}$ to $(E^x)'$, which have not been included in \cite{MimeticCGHS}
and do not conform with the mimetic correspondence.

\section{Approaches within loop quantum gravity}
\label{s:Approaches}

Given the challenging nature of implementing a consistent quantization of the
gravitational constraints and finding physically relevant solutions, different
approaches have been developed within loop quantum gravity in order to address
this problem. Since the intention is to find more tractable procedures,
some of these methods lead to potential shortcuts toward a physical solution
space. However, since they forgo a full implementation of the constraints,
their brackets, and geometrical conditions, such methods are not guaranteed to imply
viable solutions unless their consistency can be checked by independent
means.

The prime example, referred to also in \cite{MimeticCGHS}, is
deparameterization, in which one reformulates the constraints classically,
often based on special matter ingredients, and quantizes only a Hamiltonian
operator with respect to a fixed matter clock, rather than a complete
Hamiltonian constraint that could be applied with different gauge or clock
choices. In the full theory of loop quantum gravity, this viewpoint has been
espoused for instance in \cite{AQGI,DeparamQG}. A necessary consistency question is
whether this approach, which builds on a fixed gauge, can be compatible with
gauge-independent properties. Evaluations of this question are hard because
they require a partial or complete undoing of the fixed gauge or clock choice,
but they can be performed at least in spherically symmetric models.

The results of the present paper can be interpreted as ingredients of just
such an evaluation. General covariance, one of the main topics studied here,
is by definition a question about how different gauge choices can be related
to one another in a specific way that resembles coordinate transformations
applied to a metric tensor. This question cannot be addressed on a reduced
phase space or with a fixed gauge, but it is possible to ask whether reduced
or gauge-fixed models have extensions within a specific framework that are
compatible with general covariance.

The answer to this questions depends on which specific framework is used to
formulate possible extensions of a deparameterized or gauge-fixed model. In
our analysis, we used the framework of models of loop quantum gravity, which
can be defined as modified spherically symmetric theories in which the
classical quadratic dependence on the gravitational momenta has been replaced
by non-polynomial, and usually trigonometric functions. This definition is
motivated by the ubiquitous and, in fact, eponymous use of holonomies in loop
quantizations. It aims to model intricate quantum constructions as performed
for instance in
\cite{QSDI,ThreeDeform,TwoPlusOneDef,TwoPlusOneDef2,AnoFreeWeakDiff,AnoFreeWeak,OffShell,SphSymmOp,ConstraintsG}
in a simpler setting of modified classical expressions, using some of the same
functional ingredients in definitions of the constraints. As we have seen,
this approach is subject to strong covariance conditions that, in particular,
rule out the possibility of embedding the modified constraints of
\cite{MimeticCGHS} in this framework.

The covariance claims made in \cite{MimeticCGHS} were based on a different
strategy: Terms inspired by loop quantum gravity were used only in the
preferred gauge based on deparameterization of the canonical theory. The
corresponding Hamiltonian can be related to the Hamiltonian of a mimetic
theory in the preferred slicing defined by spatially constant $\phi$ used as
time, $\phi=t$. Since the mimetic side of this correspondence by construction
has a covariant extension in which the gauge fixing is undone, it may be used
as a covariant theory. In this viewpoint, therefore, models of loop quantum
gravity as a framework are replaced by models of mimetic gravity. Since these
are two different frameworks, it may well be that a covariant extension exists
only in one version, here using mimetic gravity. However, since mimetic
theories then replace models of loop quantum gravity, it remains to be
analyzed to what degree these extensions can be interpreted as related to loop
quantum gravity in slicings or gauges other than the preferred one. Another
important question is whether such an extension is unique, which would be
required for an equivalence claim.

These questions can be addressed in a straightforward manner because they
merely requires a Hamiltonian analysis of spherically symmetric mimetic
gravity for unrestricted scalar fields. It should not come as a surprise that
the new equations are much longer than the neat expressions found in the
preferred slicing because they also contain terms with $\phi'$, $\phi''$ as
well as $\dot{\phi}$ and $\ddot{\phi}$. The mimetic condition no longer
implies a direct relationship between $N$ and $\dot{\phi}$ but instead amounts
to a condition on $N$, $M$ as well as $\dot{\phi}$ and $\phi'$.  Detailed
properties of these equations can then be used to discuss whether they may be
motivated by loop quantum gravity. If this is not the case, the appearance of
loop-like Hamiltonians in the preferred slicing would merely be a coincidence
but could not be considered a generic feature of the covariant extension.

This discussion requires the following expressions: The mimetic dependence of
the action on the scalar field as used in \cite{MimeticCGHS} refers to two
terms,
\begin{eqnarray}
    g_{(2)}^{\bar{\mu}\bar{\nu}} \nabla_{\bar{\mu}} \nabla_{\bar{\nu}} \phi
    &=&
    g_{(2)}^{\bar{\mu}\bar{\nu}} \left(\partial_{\bar{\mu}} \partial_{\bar{\nu}} \phi - 
\Gamma^{\bar{\alpha}}_{\bar{\mu}\bar{\nu}} \partial_{\bar{\alpha}} \phi \right)
    \\
    &=&
    - \frac{\dot{\phi} - M \phi'}{2 N^2 E^x} \left( 2 E^x \frac{\dot{E}^\varphi}{E^\varphi} - \dot{E}^x \right)
    \nonumber\\
    &&
    - \frac{\dot{\phi}}{2 N^2 E^x} \left( M (E^x)'
    - \frac{2 E^x}{E^\varphi} (M E^\varphi)'
    + \frac{2 E^x}{N} \left( M N'
    - \dot{N}\right) \right)
    \nonumber\\
    &&
    - \frac{\ddot{\phi} - \left(N^2 E^x/(E^\varphi)^2 - M^2\right) \phi'' - 2 M \dot{\phi}'}{N^2}
    \nonumber\\
    &&
    + \frac{\phi'}{2 N^2 E^x} \Bigg(
    2 E^x \left( \dot{M} - M \frac{\dot{N}}{N}\right)
    + 2 E^x \frac{N'}{N} \left( \frac{E^x}{(E^\varphi)^2} N^2+ M^2\right)
    \nonumber\\
    &&
    - 2 E^x \frac{(E^\varphi)'}{E^\varphi} \left(\frac{E^x}{(E^\varphi)^2} N^2+ M^2\right)
    + (E^x)' \left( \frac{E^x}{(E^\varphi)^2} N^2+M^2\right)-4 E^x M M'
    \Bigg)\nonumber
\end{eqnarray}
and
\begin{equation} \label{fullY}
    Y= - g_{(2)}^{\bar{\mu}\bar{\nu}} \frac{\partial_{\bar{\mu}} E^x}{2 E^x}
    \partial_{\bar{\nu}} \phi = \frac{\dot{E}^x
      \dot{\phi}-M\left((E^x)'\dot{\phi} + \dot{E}^x \phi' \right) - \left(
        N^2 E^x/(E^\varphi)^2 - M^2\right) (E^x)' \phi'}{2N^2 E^x}\,,
\end{equation}
where $g_{(2)}$ is used to denote the 2-dimensional metric of the $t-x$
hypersurface and the barred indices correspond to only the $(t,x)$-components.
The latter expression is directly used in \cite{MimeticCGHS} as one of the
independent expressions in the function $L$ (or $L'$ after reduction to
spherical symmetry), called $Y$ in the spherically symmetric reduction. The
second function in $L'$ is given by
\begin{eqnarray}
  X&=&
  - g_{(2)}^{\bar{\mu}\bar{\nu}} \nabla_{\bar{\mu}} \nabla_{\bar{\nu}} \phi
  - g_{(2)}^{\bar{\mu}\bar{\nu}} \frac{\partial_{\bar{\mu}} E^x}{2 E^x} \partial_{\bar{\nu}} \phi
  \nonumber\\
    &=&
    \frac{\dot{\phi} - M \phi'}{N^2} \frac{\dot{E}^\varphi}{E^\varphi}
    + \frac{\dot{\phi}}{N^2} \left( \frac{\dot{E}^x}{2 E^x}
    - \frac{(M E^\varphi)'}{E^\varphi}
    + \frac{M N'
    - \dot{N}}{N} \right)
    \nonumber\\
    &&
    + \frac{\ddot{\phi} - \left(N^2 E^x / (E^\varphi)^2 - M^2\right) \phi'' - 2 M \dot{\phi}'}{N^2}
    \nonumber\\
    &&
    - \frac{\phi'}{2 N^2 E^x} \Bigg(
    2 E^x \left( \dot{M} - M \frac{\dot{N}}{N}\right)
    + 2 E^x \frac{N'}{N} \left( \frac{E^x}{(E^\varphi)^2} N^2+ M^2\right)
    \nonumber\\
    &&
    - 2 E^x \frac{(E^\varphi)'}{E^\varphi} \left(\frac{E^x}{(E^\varphi)^2} N^2+ M^2\right)
    + (E^x)' \frac{2 E^x}{(E^\varphi)^2} N^2-4 E^x M M'
    \Bigg)\,.
  \label{fullX}
\end{eqnarray}

It is easy to see that (\ref{fullX}) and (\ref{fullY}) reduce to (\ref{XY}) in
the preferred slicing of spatially constant $\phi$. The mimetic condition can
then be used to eliminate $\dot{\phi}$, and $X$ and $Y$ are directly related
to the extrinsic-curvature components of a spherically symmetric metric,
independent of the scalar field.  In other slicings,
however, the full $X$ and $Y$ remain scalar dependent, and the mimetic
condition does not seem to imply noteworthy simplifications. As a consequence,
$X$ and $Y$ in the mimetic function $L$ no longer equal components of
extrinsic curvature. Therefore, there is no choice of $L$ in such a slicing
that would turn the classical quadratic dependence on the gravitational
momenta into a dependence through trigonometric functions, as seen for
instance in the reduced action
\begin{eqnarray}
  S[g,\phi,\lambda] &=&
  \frac{1}{2G} \int{\rm d}t{\rm d}x N E^{\varphi}\sqrt{E^x}\Bigg(-2\left(\frac{\dot{E}^{\varphi}-(ME^{\varphi})'}{NE^{\varphi}}\right)\left(\frac{\dot{E}^x-M(E^x)'}{2NE^x}\right)
    \nonumber\\
    &&\qquad
  +\left(\frac{\dot{E}^x-M(E^x)'}{2NE^x}\right)^2
    +\frac{1}{2}R^{(3)}
    \nonumber\\
    &&\qquad
    + \frac{1}{2} \left( L' (X,Y)+
    \lambda\left( g_{(2)}^{\bar{\mu}\bar{\nu}} (\nabla_{\bar{\mu}}\phi)
       (\nabla_{\bar{\nu}} \phi) + 1 \right) \right) 
    \Bigg)
    \label{eq:Mimetic action-spherical}
\end{eqnarray}
for a non-uniform slicing.

The methods of loop quantum gravity can then be
used to suggest modifications of the Hamiltonian only in a preferred slicing,
which is tantamount to saying that in this viewpoint, loop quantum gravity is
defined only on a preferred slicing of space-time. On other slices, the
paradigm of loop quantum gravity is replaced by theories of mimetic
gravity. Moreover, this definition of loop quantum gravity can only be made in
the presence of a scalar field. There would be no vacuum theory of loop
quantum gravity.

An independent concern is that the mimetic extension is far from being unique,
starting with specific modifications of the deparameterized canonical
theory. In \cite{MimeticCGHS}, a unique mimetic extension was derived for
mimetic gravity in which $L$ depends only on two local scalar invariants,
$\nabla^{\mu}\nabla_{\mu}\phi$ and
$(\nabla_{\mu}\nabla_{\nu}\phi)(\nabla^{\mu}\nabla^{\nu}\phi)$, reduced to $X$
and $Y$ in spherical symmetry. Mimetic gravity can be formulated with an
infinite number of higher-order invariants with different contractions of $n$
factors $\nabla_{\mu}\phi$, or $\nabla_{\bar{\mu}}\phi$ as well as
$\nabla_{\bar{\nu}}E^x$ in a restriction to spherical symmetry. Any term that
reduces to at most one derivative of $E^x$ in spherical symmetry could play
the role of $Y$ in the preferred slicing, but it would imply an inequivalent
mimetic extension of the deparameterized canonical theory. Moreover, the
mimetic condition is not strictly required for these constructions because it
is trivialized by the gauge choice $\phi=t$, and can therefore be replaced by
this choice and then generalized to non-mimetic theories, such as those of
Horndeski type. Since such covariant extensions of deparameterized canonical
theories are not uniquely defined by the deparameterized theory, they cannot
be considered equivalent descriptions.

\section{Conclusions}

We have demonstrated that the constructions given in \cite{MimeticCGHS} fail
to give a faithful description of models of loop quantum gravity. By
restricting the covariant theory of mimetic gravity to a preferred slicing,
the resulting models not only lose reliable access to covariance properties
within the paradigm of loop quantum gravity, they are also non-unique and
obscure important dynamical features of the related theories. If a theory is
known to be covariant, it may well be analyzed in a preferred slicing without
losing any physical information. However, if a theory such as loop quantum
gravity, whose covariance status is unclear, is related to a canonical theory
in a preferred slicing, it is impossible to draw conclusions about any
equivalence between the theories. This statement is clearly demonstrated by
the example of Ho\v{r}ava--Lifshitz gravity. In this example,  a specific form of the
4-dimensional action principle replaces the framework of models of loop
quantum gravity, but conceptually it gives rise to the same equivalence question. 

Full equivalence between two general frameworks in which actions or
Hamiltonians are specified by independent principles, which defines the
desired kind of equivalence in the present context, can only be obtained if a
comparison of equations in a preferred slicing is accompanied by a detailed
analysis of gauge transformations. If gauge transformations on both sides of
the correspondence are equivalent and equations of motion agree in a preferred
gauge or slicing, the theories are equivalent in their dynamics as well as
symmetry properties. For instance, models of emergent modified gravity can be
related to mimetic gravity because they are covariant by construction, but the
correspondence then requires using the correct emergent space-time metric that
must be derived from gauge properties. (In spherical symmetry, the radial
component of the emergent metric need not equal $(E^{\varphi})^2/E^x$.) Models
of loop quantum gravity, by contrast, cannot be equivalent to covariant
theories unless additional terms (such as $K_{\varphi}'$) are included as
suggested by emergent modified gravity.

The mimetic correspondence proposed in
\cite{MimeticCGHS} does not imply a strong equivalence because there is only
one full, non-gauge fixed theory in this case, given by a mimetic action. The
proposal simply defines the non-gauge fixed version of a deparameterized model
of loop quantum as being the same as the constructed mimetic theory. This
procedure constitutes a definition but not the derivation of an
equivalence. Moreover, as demonstrated here, the resulting theory is far from
being unique even if a specific set of loop modifications is used in the
deparameterized theory.

The constructions in \cite{MimeticCGHS} and other examples in loop quantum
gravity attempt to evade a detailed discussion of anomaly freedom, covariance,
and gauge transformations by using the canonical theory in deparameterized
form, schematically replacing constraint equations $C=0$ with equations $C=P$
where $P$ represents terms linear in the momenta of matter fields.  It is then
argued that all gauge transformations of the complicated gravitational terms
in $C$ can be replaced by simple gauge transformations generated by
expressions linear in momenta. However, terms linear in momenta do not
reproduce the brackets of hypersurface deformations required for general
covariance or slicing independence in a canonical theory. These brackets can
be reproduced faithfully only by the full $C-P$ (if matter terms are desired).
Consistency conditions then rule out arbitrary modifications such as
spherically symmetric contributions to the Hamiltonian constraint that are not
linear in $K_x$. As a corollary, the failed correspondence analyzed here
therefore demonstrates that deparameterization is not a reliable way to
construct generally covariant modified theories of gravity.

\section*{Acknowledgements}

The authors are grateful to Muxin Han for discussions.
This work was supported in part by NSF grant PHY-2206591.

%\bibliographystyle{../preprint}
%\bibliography{../Bib/QuantGra}

\end{document}